# Emission of Terahertz Waves from Stacks of Intrinsic Josephson Junctions

K.E. Gray, L. Ozyuzer, A.E. Koshelev, C. Kurter, K. Kadowaki, T. Yamamoto, H. Minami, H. Yamaguchi, M. Tachiki, W.-K. Kwok, and U. Welp

*Abstract*—By patterning mesoscopic crystals of $Bi_2Sr_2CaCu_2O_8$ (BSCCO) into electromagnetic resonators the oscillations of a large number of intrinsic Josephson junctions can be synchronized into a macroscopic coherent state accompanied by the emission of strong continuous wave THz-radiation. The temperature dependence of the emission is governed by the interplay of self-heating in the resonator and by re-trapping of intrinsic Josephson junctions which can yield a strongly non-monotonic temperature dependence of the emission power. Furthermore, proper shaping of the resonators yields THz-sources with voltage-tunable emission frequencies.

## I. INTRODUCTION

ELECTROMAGNETIC waves at THz-frequencies hold great promise for noninvasive sensing, imaging and spectroscopy across the physical, medical, pharmaceutical and biological sciences as well as for new applications in security and surveillance settings, and high-bandwidth communications [1]-[3]. However, progress has been limited by the lack of compact solid-state sources of powerful continuous-wave THz-radiation, a situation commonly summarized as the 'THz-gap'. The Josephson effect occurring between superconducting electrodes that are separated by a thin barrier layer constitutes a natural converter for transforming a dc-voltage into high-frequency electromagnetic (em) fields, with 1 mV corresponding to 0.483 THz. Thus Josephson junctions [4], [5] and arrays of Josephson junctions [6], [7] are potential sources of high-frequency em radiation. Powers of 10 μW at frequencies up to 500 GHz were delivered by an array of 500 $Nb/AlO_x/Nb$ junctions to a detector junction [8] whereas an array of 1968 junctions generated 160 μW at 240 GHz [9]. Local oscillators for use in superconducting receivers based on flux flow oscillators delivered several μW at frequencies around 450 GHz [10]. The discovery of the intrinsic Josephson junctions [11] in the crystal structure of the layered high-temperature superconductor $Bi_2Sr_2CaCu_2O_8$ (BSCCO) enables the fabrication of 1D arrays in the form of stacks or so-called mesas sculpted from BSCCO crystals that contain a very large number of identical junctions. These are expected to allow the fabrication of novel sources of intense, coherent THz-radiation [12]-[14] that operate at frequencies well exceeding those obtainable with Nb-technology due to the large superconducting gap energy of BSCCO and at temperatures close to 77 K. However, the requirement of achieving synchronization of the high-frequency oscillations of all the junctions in the stack has emerged as a major hurdle.

Recently, we have demonstrated that stacks of intrinsic Josephson junctions in BSCCO can be induced to emit coherent continuous-wave radiation in the THz-frequency range [15]. These samples were designed in such a way that an electromagnetic cavity resonance synchronizes a large number of intrinsic junctions into a macroscopic coherent state enabling emission powers of 0.5 μW and frequencies up to 0.85 THz. More recently, radiation powers of 5 μW and emission frequencies – at the fourth harmonic – of up to 2.5 THz have been obtained with BSCCO [16]. Recent modeling suggests that the micron-sized samples used in these studies hold the potential of emitting at powers of ~1 mW under optimized conditions [17], [18].

Here we present new results on the temperature and voltage dependence of the emission from BSCCO mesas.

Manuscript received 19 August 2008. This work was supported by the US-Department of Energy, Basic Energy Sciences, under Contract No. DE-AC02-06CH11357, by JST (Japan Science and Technology Agency) CREST project, by the JSPS (Japan Society for the Promotion of Science) CTC program, by the Grant-in Aid for Scientific Research (A) under the Ministry of Education, Culture, Sports, Science and Technology (MEXT) of Japan, and by the Turkish TUBITAK under Project No. 106T053.

K. E. Gray, A. E. Koshelev, C. Kurter, W.-K. Kwok, and U. Welp are with the Materials Science Division, Argonne National Laboratory, Argonne, IL 60439 USA (K. E. Gray: 630-252-5525; fax: 630-252-4748; e-mail: kengray@anl.gov).

K. Kadowaki, T. Yamamoto, H. Minami, and H. Yamaguchi, are with the Institute of Materials Science, University of Tsukuba, 1-1-1 Tennodai, Ibaraki-ken 305-8577, Japan (e-mail: kadowaki@ims.tsukuba.ac.jp).

M. Tachiki is with the Graduate School of Frontier Sciences, The University of Tokyo, 5-1-5 Kashiwanoha, Kashiwa 277-8568, Japan (e-mail: tachiki@k.u-tokyo.ac.jp).

## II. EXPERIMENT AND RESULTS

The operation of the THz-source is based on the propagation of electromagnetic waves – the Josephson plasma modes [19]-[21] – inside the layered BSCCO structure. The mode with the highest velocity, $c_0/n$, where $n$ is the far-IR refractive index of BSCCO for c-axis polarized waves, is the in-phase mode (all junctions oscillate in-phase) and the mode with the lowest velocity is the anti-phase mode in which neighboring junctions oscillate out-of-phase. Only the in-phase mode produces noticeable emission. In mesas with lateral dimensions that are smaller than the propagation distance of the Josephson plasma modes, multiple reflections at the side faces of the mesa create a standing wave pattern and Fabry-Perot type cavity resonances, or Fiske resonances. For our samples, we estimate a propagation distance of $l = \sqrt{\beta_C}\lambda_c \sim 1$ cm, where $\beta_C \sim 10^4$ is the McCumber parameter and $\lambda_c \sim 100$ μm is the c-axis magnetic penetration depth. A resonance condition occurs whenever the Josephson frequency, $f_{Jos}$, equals the cavity frequency, $f_{cav}$: $f_{cav} = c_0/2nw = f_{Jos} = V/N\Phi_0$. Here, w is the width of the cavity, V is the voltage applied across the junction stack, $\Phi_0$ is magnetic flux quantum, and N is the number of active junctions in the stack. As the resonance condition is approached, for example by scanning the applied voltage, energy is pumped into the cavity resonance. As the cavity field builds up it entrains the oscillations of more junctions leading to a further increase of the cavity field until (almost) all the junctions are synchronized into a coherent in-phase oscillation accompanied by a large cavity field. This process resembles the emergence of coherence in a laser cavity. Recent low-temperature scanning laser microscopy of the electric field distribution in BSCCO mesas [22] supports the model of the formation of a resonant cavity mode. Through the electromagnetic boundary conditions at the end-faces of the cavity, THz-radiation is transmitted into free-space. Furthermore, the resonance condition implies that the emission frequency can be controlled by the mesa width according to f ~ 1/w, and this relation has been verified in our previous experiments [18]. The energy for the radiation is supplied by the Josephson oscillations. Several mechanisms of transferring the energy from the Josephson oscillations into the cavity resonance have recently been analyzed theoretically [17], [18], [23]-[26].

The BSCCO resonators used in the present study are mesas fabricated using Argon-ion milling and photolithography techniques from cleaved BSCCO crystals that were grown by a traveling floating zone method [27]. The dimensions of the mesas are 300 μm in length, w = 100, 80, 60 and 40 μm in width, and about 1 μm in height. Due to the manufacturing process the mesas are narrower at the top than at the base as discussed in more detail below. A schematic layout is shown in Fig. 1. When a current passed down the mesa is adjusted in such a way that the corresponding voltage fulfils the above resonance condition, THz-radiation is emitted from the long side faces of the mesa. The superconducting transition temperatures of 77 K to 80 K are the result of underdoping. The radiation power was measured using a liquid-helium-cooled Si-bolometer and the emission spectra were obtained using a FT-IR spectrometer enabling measurements between 2

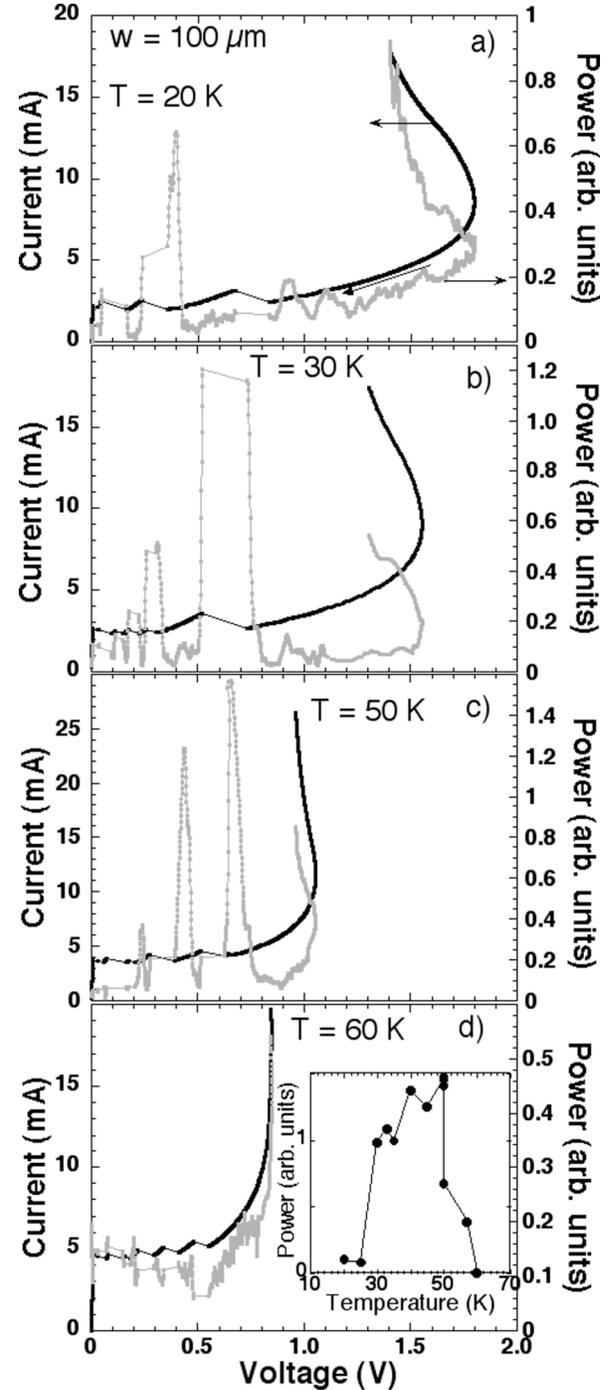

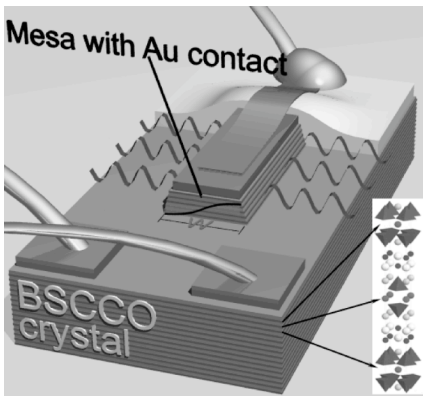

Fig. 1. Schematic of a BSCCO resonator. As current passes down the mesa, it excites the fundamental cavity resonance on the width w of the mesa, illustrated by the black half-wave. THz-radiation is emitted from the long side-faces of the mesa as shown by the grey waves. The inset illustrates the layered crystal structure of BSCCO.

Fig. 2. Temperature dependence of the current-voltage curve (dark lines, left y-axis) and of the emission power (grey lines, right y-axis) measured on decreasing bias current on a 100-μm wide mesa. A non-monotonous temperature dependence of the emission power (inset of panel d)) results from the interplay of self-heating of the mesa indicated by the back-bending of the IV-curve at high bias and of re-trapping seen as jumps in the IV-curve.

and 650 cm$^{-1}$ with a resolution of 0.25 cm$^{-1}$ = 7.5 GHz.

Figure 2 shows the current-voltage (IV) characteristics and the radiation power of a 100-μm wide mesa recorded simultaneously while decreasing the bias current at various sample temperatures. This mesa emits at a frequency of 0.36 THz. Previously we have shown that emission is observed only when the bias is decreased along the fully resistive McCumber branch [15]. The back bending of the IV-curve at high currents is due to self-heating effects as indicated by the appearance of unpolarized black-body radiation [15]. Upon decreasing the bias, emission peaks, polarized perpendicular to the CuO$_2$-planes, occur at specific voltages (see Fig. 2c) signaling Josephson radiation [15]. The emission peak at the highest voltage corresponds to the participation of essentially all the junctions in the stack. On further decreasing bias a certain number of junctions re-trap from the resistive into the non-radiating superconducting state, seen as thin-line diagonal jumps in the IV-curves, and a second emission peak involving fewer active junctions may arise (see below). The temperature evolution of the emission (Figs. 2a-d and inset) is mainly due to the interplay of two phenomena, the higher likelihood of re-trapping events at low temperatures and the approach to $T_c$ due to self-heating at high temperatures. At 20 K, the highest voltage emission peak is absent because a re-trapping event bypasses the resonance voltage for the full stack. At 30 K, this higher voltage peak appears, although it appears to be cut short by re-trapping. At 50 K two consecutive full emission peaks are observed without re-trapping. Finally, at 60 K self-heating arises at such a low voltage that it interferes with the emission, and the mesa is locally driven to the normal state. The inset of Fig. 2 summarizes this behavior of the highest voltage emission peak that is observed on many – but not all – mesas; see also [16].

The loss of radiation at low temperatures by re-trapping can be understood qualitatively. The dynamics of a Josephson junction can formally be mapped onto the motion of a particle in a tilted washboard potential [28] where the overall tilt is proportional to the drive current and the depth of the relief is given by Josephson critical current. Re-trapping corresponds to the particle becoming localized in one potential well. Upon decreasing temperature, the critical current increases whereas the (applied) quasiparticle current is relatively unchanged (see Fig. 2). Thus it is the larger depth of the potential wells that enhance the probability of re-trapping. Re-trapping is a sample-dependent statistical phenomenon [29]-[35] and in large junction arrays it needs further clarification.

The IV-curve and emission power for an 80-μm wide mesa is shown in Fig. 3a. Three emission peaks are resolved as the voltage decreases. The FT-IR emission spectra acquired at these peak voltages are displayed in Fig. 3b. They directly demonstrate that the emission frequency is virtually the same for each of these peaks. On decreasing the bias from the fully resistive state the emission power builds up as the Josephson frequency comes in resonance with the cavity (#1). During a jump in the IV-curve some junctions switch to the superconducting state (white regions in the inset of Fig. 3b) while the voltage per remaining resistive junction jumps up. Consequently, the Josephson frequency increases and falls out of resonance with the cavity mode until a further decrease in bias regains the resonance. Thus the second emission peak (#2) involves a smaller number of active junctions but is at the same frequency.

The data in Fig. 2c reveal an appreciable width in voltage of the emission peaks that might indicate a range of resonance frequencies. This is shown in greater detail in Fig. 4 that displays emission peaks and power spectra of 60-μm and 100-μm mesas. Emission in Fig. 4a occurs over a voltage range of ~50 mV and the FT-IR spectra corresponding to the voltages labeled 1 though 12 are shown in Fig. 4b. For each voltage we observe a resolution-limited emission line-width with an intensity that tracks the over–all emission power in Fig. 4a. Thus the emission frequency can be tuned by scanning the voltage through the peak by about 7%. We note that the voltage can be scanned in a reversible manner as evidenced by spectrum #12, which was acquired after increasing the voltage from level #11. This spectrum coincides with spectrum #9 taken while decreasing the voltage. An irreversible re-trapping event occurs near 0.72 V that is well away from this emission band. The power spectra of the 100-μm mesa, shown in Fig. 4c, display (see also Fig. 2b,c) a similar trend as those in Fig. 4b. However, the range of tunability is significantly smaller. We explain these data by realizing that the mesas have trapezoidal cross sections (Fig. 5) with sloping angles of the long side faces that are as low as 20 to 30 deg. Whereas in an ideal cavity a sharp resonance at a well-defined voltage is expected, for the trapezoidal cross section a distribution of resonance frequencies arises corresponding to the varying width as shown schematically in Fig. 5 (middle, right). On decreasing voltage the resonance condition is first met at the narrow top of the mesa and subsequently at decreasing heights within the mesa. Synchronization of oscillations in different junctions for Josephson-junctions arrays coupled via external cavities is theoretically investigated using different versions of the Kuramoto model [36], [37]. In our case the coupling of the junctions is established through the resonant cavity field. Recent numerical simulations of the dynamics of junction stacks based on the Lawrence-Doniach model support this scenario. The stronger this coupling the more junctions will be synchronized as represented by the dark band in Fig. 5, middle. Thus, each line in the power spectrum (Fig. 4b) corresponds to a band at a location that fulfils an 'average' resonance condition and with a width given by the observed line intensity. We note that the line intensities are appreciable indicating that a large number of junctions synchronize at a given frequency. At low voltages this band leaves the mesa at the bottom, and emission ceases. Since the sloping angles are roughly the same for most mesas the relative variation in the width is much larger for the narrower mesa, its range of resonance frequencies will accordingly be larger, thereby explaining the large difference in the range of tunability seen in Fig. 4b and 4c.

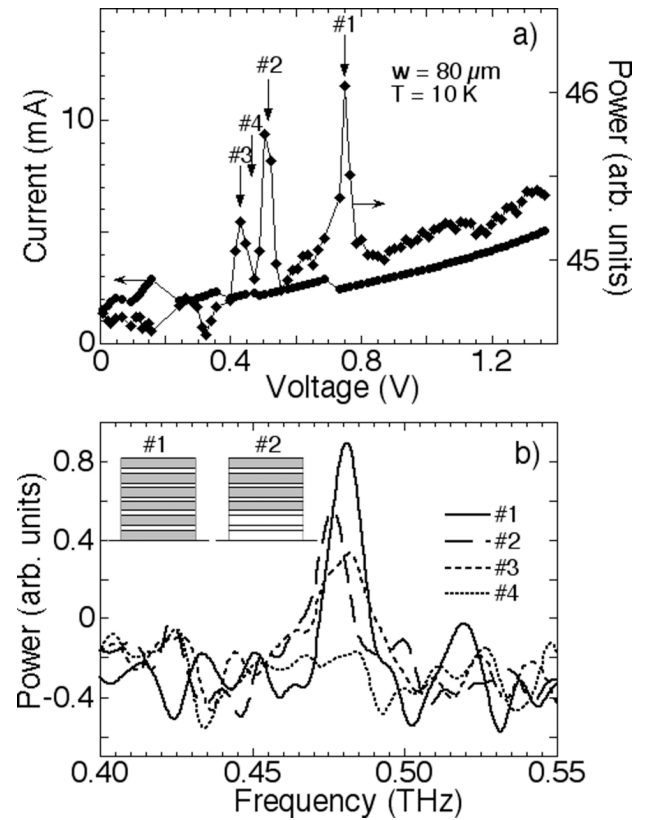

Fig. 3. (a) Voltage dependence of the current and of the emission power measured on a 80-μm mesa. Three successive emission peaks corresponding to different numbers of active junctions are resolved. (b) Emission spectra at voltages corresponding to the three peaks in (a). The fourth spectrum is the baseline. These data show directly that the emission peaks seen in the voltage dependence have the emission frequency. The inset illustrates a fully resistive junction stack (#1) and a stack in which some junctions re-trapped into the superconducting state (white) (#2), respectively.

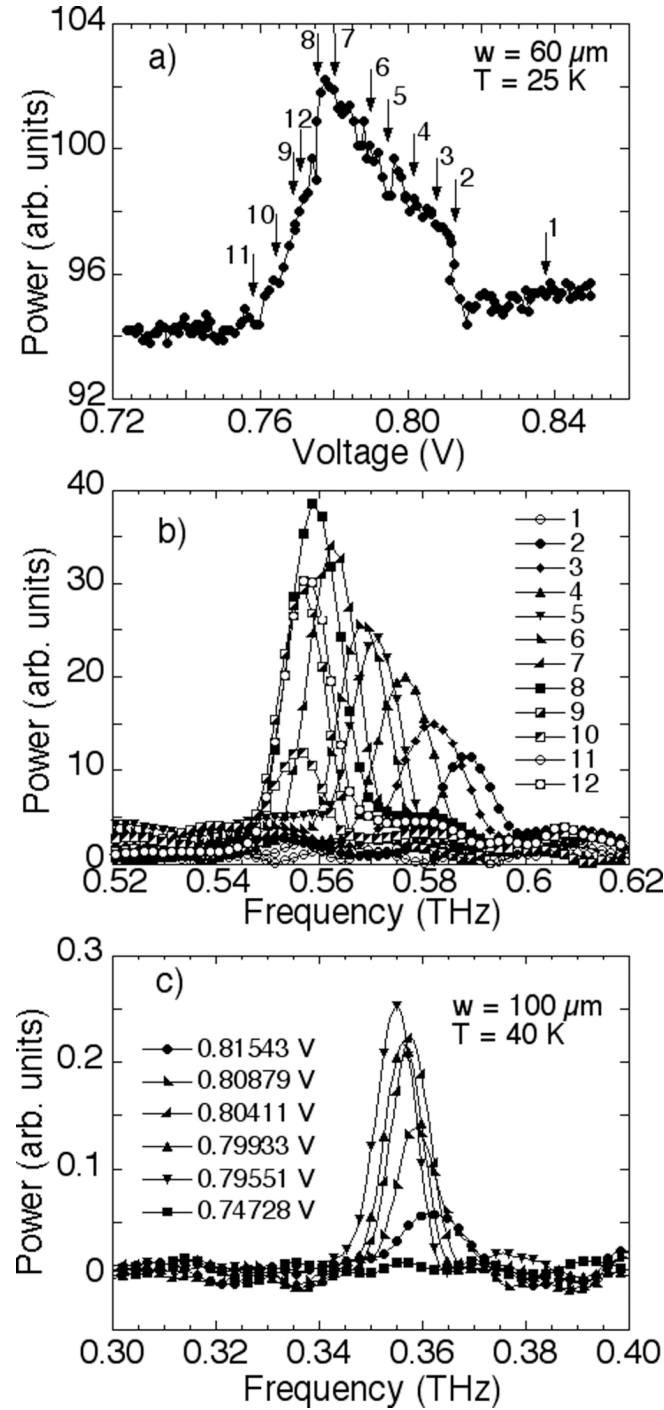

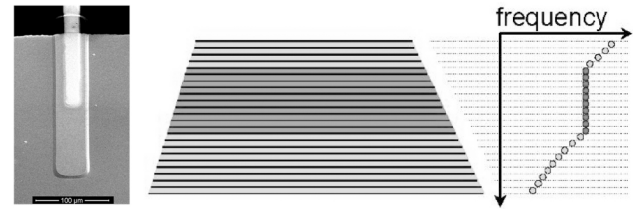

Fig. 5. Electron microscopy image of a 40-μm mesa (left) revealing the trapezoidal mesa cross-section. The black contrast at the top of the image is the $CaF_2$-insulating layer [12]. Schematic of the cross section (middle). The dark band represents junctions that are synchronized to a single frequency. Schematic of the z-dependence of the frequency (right).

Fig. 4. Voltage dependence of the emission. a) Emission peak of a 60-μm mesa. The numbers label the voltages at which the power spectra shown in Fig. 4b) have been taken. c) Power spectra of the 100-μm mesa at various voltages (other data on this mesa are shown in Fig. 2).

### III. CONCLUSIONS

We find that the temperature dependence of the THz wave emission from BSCCO mesas is governed by the interplay of self-heating in the resonator at high temperatures and be re-trapping of intrinsic junctions at low temperatures resulting in a strongly non-monotonic temperature dependence of the emission power. Furthermore, proper shaping of the resonators yields THz-sources with voltage-tunable emission frequencies.

### ACKNOWLEDGMENT

We thank A. Imre at the Center for Nanoscale Materials, Argonne National Laboratory, for help with the electron microscopy and FIB work.

### REFERENCES

[1] M. Tonouchi, "Cutting-edge terahertz technology," *Nature Photonics*, vol. 1, pp. 97-105, Feb. 2007.

[2] For a series of recent reviews: *Proceedings of the IEEE*, vol. 95, nr. 8, pp. 1515-1704, Aug. 2007.
[3] M. Lee and M. C. Wanke, "Searching for a Solid-State THz-Technology," *Science*, vol. 316, pp. 64-65, Apr. 2007.
[4] I. M. Dmitrenko, I. K. Yanson, and V. M. Svistunov, "Experimental observation of the tunnel effect of Cooper pairs with the emission of photons," *Pis'ma ZhETF*, vol. 65, pp. 650-652, 1965.
[5] D. N. Langenberg, D. J. Scalapino, B. N. Taylor, R. E. Eck, "Investigation of Microwave Radiation Emitted by Josephson Junctions," *Phys. Rev. Lett.*, vol. 15, pp. 294-297, 1965.
[6] M. Darula, T. Doderer, and S. Beuven, "Millimetre and sub-mm wavelength radiation sources based on discrete Josephson junction arrays." *Supercond. Sci. Technol.*, vol. 12, pp. R1-R26, 1999.
[7] A. K. Jain, K. K. Likharev, J. E. Lukens, and J. E. Sauvageau, "Mutual phase-locking in Josephson junction arrays." *Phys. Rep.*, vol. 109, pp. 309-426, 1984.
[8] S. Han, B. Bi, W. Zhang, and J. E. Lukens, "Demonstration of Josephson effect submillimeter wave sources with increased power," *Appl. Phys. Lett.*, vol 64, pp. 1424-1426, March 1994.
[9] P. A. A. Booi and S. P. Benz, "High power generation with distributed Josephson-junction arrays," *Appl. Phys. Lett.* Vol 68, pp. 3799-3801, June 1996.
[10] V. P. Koshelets and S. V. Shitov, "Integrated superconducting receivers," *Supercond. Sci. Technol.* Vol 13, pp. R53-R69, 2000.
[11] R. Kleiner, F. Steinmeyer, G. Kunkel, and P. Müller, "Intrinsic Josephson effect in $Bi_2Sr_2CaCu_2O_8$ single crystals." *Phys. Rev. Lett.*, vol 68, pp. 2394-2397, 1992.
[12] T. Koyama and M. Tachiki, "Plasma excitation by vortex flow," *Solid State Commun.*, vol. 96, pp. 367-371, 1995.
[13] M. Tachiki, M. Iizuka, K. Minami, S. Tejima, and H. Nakamura, "Emission of continuous coherent terahertz waves with tunable frequency by intrinsic Josephson junctions," *Phys. Rev. B*, vol. 71, pp. 134515-1 – 134515-5, Apr. 2005.
[14] S. Savel'ev, A. Rakhmanov, and F. Nori, "Using Josephson Vortex Lattices to Control Terahertz Radiation: Tunable Transparency and Terahertz Photonic Crystals," *Phys. Rev. Lett.*, vol. 94, pp. 157004-1 - 157004-4, Apr. 2005.
[15] L. Ozyuzer, A. E. Koshelev, C. Kurter, N. Gopalsami, Q. Li, M. Tachiki, K. Kadowaki, T. Yamamoto, H. Minami, H. Yamaguchi, T. Tachiki, K. E. Gray, W.-K. Kwok and U. Welp, "Emission of Coherent THz-Radiation from Superconductors," *Science*, vol. 318, pp. 1291-1293, Nov. 2007.
[16] K. Kadowaki, H. Yamaguchi, K. Kawamata, T. Yamamoto, H. Minami, I. Kakeya, U. Welp, L. Ozyuzer, A. E. Koshelev, C. Kurter, K. E. Gray, and W.-K. Kwok, "Direct observation of terahertz electromagnetic waves emitted from intrinsic Josephson junctions in single crystalline $Bi_2Sr_2CaCu_2O_{8+\delta}$,", *Physica C*, vol. 468, pp. 634-639, March 2008.
[17] A. E. Koshelev and L. N. Bulaevskii, "Resonant electromagnetic emission from intrinsic Josephson-junction stacks with laterally modulated Josephson critical current," *Phys. Rev. B*, vol. 77, pp. 014530-1 – 014530-15, Jan. 2008.
[18] S. Lin, X. Hu and M. Tachiki, "Computer simulations of terahertz emission from intrinsic Josephson junctions of high-$T_c$ superconductors," *Phys. Rev. B*, vol. 77, pp. 014507-1 – 014507-5, Jan. 2008.
[19] R. Kleiner,"Two-dimensional resonant modes in stacked Josephson junctions," *Phys. Rev. B*, vol. 50, pp. 6919-6923, May 1994.
[20] L. N. Bulaevskii, M. Zamora, D. Baeriswyl, H. Beck, and J. R. Clem, "Time-dependent equations for phase differences and a collective mode in Josephson-coupled layered superconductors," *Phys. Rev. B*, vol. 50, pp. 12831-12834, Nov. 1994.
[21] N. F. Pedersen, and S. Sakai, "Josephson plasma resonance in superconducting multilayers," *Phys. Rev. B* **58**, 2820-2826, Aug. 1998.
[22] H. B. Wang, S. Guenon, J. Yuan, A. Iishi, S. Arisawa, T. Hatano, T. Yamashita, D. Koelle, and R. Kleiner, "Hot spots and waves in $Bi_2Sr_2CaCu_2O_{8+\delta}$, intrinsic Josephson junction stacks – a study by Low Temperature Scanning Laser Microscopy," arXiv:cond-mat/0807.2749 (July 2008).
[23] S. Lin and X. Hu, "Possible Dynamic States in Inductively Coupled Intrinsic Josephson Junctions of Layered High-$T_c$ Superconductors" *Phys. Rev. Lett.*, vol. 100, pp. 247006-1 – 247006-4, Jun. 2008.
[24] A. E. Koshelev, "Alternating dynamic state in intrinsic Josephson-junction stacks self-generated by internal resonance", arXiv:0804.0146.
[25] M. Tachiki, "Emission of terahertz electromagnetic waves by an external current in intrinsic Josephson junctions," *Physica C*, vol. 468, pp. 631-633, March 2008.
[26] R. Klemm and K. Kadowaki, "Angular dependence of the radiation power of a Josephson STAR-emitter," arXiv:0807.3082 (July 2008).
[27] T. Mochiku, and K. Kadowaki, "Preparation of $Bi_2Sr_2(Ca,Y)Cu_2O_{8+d}$ single crystals by TSFZ method," *Trans. Mat. Res. Soc. Jpn.*, vol. 19A, pp. 349-352, 1994.
[28] M. Tinkham, "*Introduction to Superconductivity*," 2$^{nd}$ edition, New York: McGraw-Hill, 1996, ch. 6.
[29] E. Ben-Jacob, "Lifetime of oscillatory steady states", *Phys. Rev. A*, vol. 26, pp. 2805-2816, 1982.
[30] T. A. Fulton and L. N. Dunkelberger, "Lifetime of the zero-voltage state in Josephson junctions," *Phys. Rev. B*, vol. 9, pp. 4760-4768, 1974.
[31] M. Büttiger, E. P. Harris and R. Landauer, "Thermal activation in extremely underdamped Josephson junction circuits, " *Phys. Rev. B*, vol. 28, pp. 1268-1275, 1983.
[32] M. G. Castellano, G. Torrioli, F. Chiarello, C. Cosmelli, P. Carelli, "Return current in hysteretic Josephson junctions : Experimental distribution in the thermal activation regime," *Journal of Applied Physics*, vol. 86, pp. 6405-6411, Dec. 1999.
[33] M. Machida, T. Koyama and M. Tachiki, "Dynamical Breaking of Charge Neutrality in Intrinsic Josephson Junctions: Common Origin for Microwave Resonant Absorptions and Multiple-Branch Structures in the I-V Characteristics," *Phys. Rev. Lett.*, vol. 83, pp. 4618-4621, Nov. 1999.
[34] V. M. Krasnov, T. Golod, T. Bauch, P. Delsing, "Anticorrelation between temperature and fluctuations of the switching current in moderately damped Josephson junctions," *Phys. Rev. B*, vol 76, pp. 224517-1 – 224517-16, Dec. 2007.
[35] A. Irie, Yu. M. Shukrinov, and G. Oya,"Experimental manifestation of the breakpoint region in the current-voltage characteristics of intrinsic Josephson junctions", *Appl. Phys. Lett.*, vol. 93, 15210-1 – 15210-3, Oct. 2008.
[36] G. Filatrella, N. F. Pedersen, and K. Wiesenfeld, "High-Q cavity-induced synchronization in oscillator arrays," *Phys. Rev. B*, vol. 61, pp. 2513-2518, March 2000.
[37] G. Filatrella, N. F. Pedersen, and K. Wiesenfeld, "Generalized coupling in the Kuramoto model," *Phys. Rev. E* , vol. 75, 017201-1 - 017201-4, Jan. 2007.